\documentclass[letterpaper, 10 pt, conference]{IEEEtran}
\IEEEoverridecommandlockouts
\usepackage{cite}
\usepackage{amsmath,amssymb,amsfonts}
\usepackage{algorithmic}
\usepackage{graphicx}
\usepackage{textcomp}
\usepackage{xcolor}
\usepackage{bm}
\usepackage[english]{babel}
\graphicspath{{figures/}}
\newtheorem{theorem}{Theorem}

\newtheorem{lemma}{Lemma}
\newtheorem{assumption}{Assumption}

\def\BibTeX{{\rm B\kern-.05em{\sc i\kern-.025em b}\kern-.08em
    T\kern-.1667em\lower.7ex\hbox{E}\kern-.125emX}}
\usepackage[hmargin=0.75in, vmargin=0.75in]{geometry}
\usepackage{afterpage}

\begin{document}
\newgeometry{hmargin=0.75in, top=1in, bottom=0.75in}


\title{
Feasibility Guaranteed Traffic Merging Control Using Control Barrier Functions\\
	\thanks{%
			This work was supported in part by NSF under grants ECCS-1931600,
			DMS-1664644 and CNS-1645681, by ARPAE under grant DE-AR0001282, by AFOSR
			under grant FA9550-19-1-0158, by the MathWorks and by the Qatar National Research Fund, a member of the Qatar Foundation (the statements made herein are solely the responsibility of the authors.)}
			\thanks{%
		    K. Xu and C. G. Cassandras are with the
			Division of Systems Engineering and Center for Information and Systems
			Engineering, Boston University, Brookline, MA, 02446, USA \{xky, cgc\}@bu.edu}
			\thanks{%
		    W. Xiao is with the
			Computer Science and Artificial Intelligence Lab, Massachusetts Institute of Technology.  weixy@mit.edu}}
	\author{\IEEEauthorblockN{Kaiyuan Xu} \and 
		\IEEEauthorblockN{Wei Xiao} \and \IEEEauthorblockN{Christos G. Cassandras} }
	\maketitle

\maketitle
\afterpage{\afterpage{\aftergroup\restoregeometry}}

\begin{abstract}
    We consider the merging control problem for Connected and Automated Vehicles (CAVs) aiming to jointly minimize travel time and energy consumption while providing speed-dependent safety guarantees and satisfying velocity and	acceleration constraints. Applying the joint optimal control and control barrier function (OCBF) method, a controller that optimally tracks the unconstrained optimal control solution while guaranteeing the satisfaction of all constraints is efficiently obtained by transforming the optimal tracking problem into a sequence of quadratic programs (QPs). However, these QPs can become infeasible, especially under tight control bounds, thus failing to guarantee safety constraints. We solve this problem by deriving a control-dependent feasibility constraint corresponding to each CBF constraint which is added to each QP and we show that each such modified QP is guaranteed to be feasible. Extensive simulations of the merging control problem illustrate the effectiveness of this feasibility guaranteed controller.
\end{abstract}

\section{Introduction}

The performance of transportation networks critically depends on the management of traffic at conflict areas such as intersections, roundabouts and merging roadways \cite{rios2016survey}. Coordinating and controlling vehicles in such conflict areas is a challenging problem in terms of reducing congestion and energy consumption while also ensuring passenger comfort and guaranteeing safety \cite{chen2015cooperative, tideman2007review}. The emergence of Connected
and Automated Vehicles (CAVs) \cite{rios2016survey} and the development of new traffic infrastructure technologies \cite{li2013survey} provide a promising solution to this problem
through better information utilization and more precise vehicle trajectory design.

Both centralized and decentralized methods have been studied to deal with
the control and coordination of CAVs at conflict areas. Centralized
mechanisms are often used in forming platoons in merging problems \cite%
{xu2019grouping} and determining passing sequences at intersections \cite%
{xu2020bi}. These approaches tend to work better when the safety constraints
are independent of speed and they generally require significant computation
resources, especially when traffic is heavy. They are also not easily
amenable to disturbances.

Decentralized mechanisms restrict all computation to be done on board each
CAV with information sharing limited to a small number of neighbor
vehicles \cite{milanes2010automated, rios2015online, bichiou2018developing,
hult2016coordination}. Optimal control problem formulations are often used, with Model Predictive Control (MPC) techniques employed
as an alternative to account for additional constraints and to compensate
for disturbances by re-evaluating optimal actions \cite{cao2015cooperative,
mukai2017model, nor2018merging}. The objectives in such problem formulations
typically target the minimization of acceleration or the maximization of
passenger comfort (measured as the acceleration derivative or jerk). An
alternative to MPC has recently been proposed through the use of Control
Barrier Functions (CBFs) \cite{xiao2020bridging, Xiao2019}
which provide provable guarantees that safety constraints are always
satisfied.

For the merging control problem under CAV traffic, a decentralized optimal merging control framework with a complete solution is given in \cite{xiao2021decentralized}. The objective jointly minimizes
(i) the travel time of each CAV over a given road segment
from a point entering a Control Zone (CZ) to the eventual
merging point and (ii) a measure of its energy consumption.
However, the computational complexity in deriving this solution, even under simple vehicle dynamics, becomes prohibitive for real-time applications when safety constraints (e.g., preventing rear-end collisions) become active. This limitation can be overcome by the joint Optimal Control with Control Barrier Functions (OCBF) approach in \cite{xiao2020bridging}. In this approach, we first derive the solution of the optimal merging control problem when no constraints become active. Then, we solve another problem to optimally track this solution while also guaranteeing the satisfaction of all constraints using CBFs \cite{Xiao2019}. Thus, the OCBF controller aims to minimize deviations from an optimal unconstrained trajectory while ensuring that no constraint is violated. As shown in \cite{xiao2020bridging}, this also allows the use of more accurate vehicle dynamics, possibly including noise, and the presence of more complicated objective functions. The OCBF controller is derived through a sequence of Quadratic Programs (QPs) over time which are simple to solve. However, they may become infeasible when the control bounds conflict with the CBF constraints, in which case the safety constraints can no longer be guaranteed. Thus, a basic question in CBF-based methods is: how can we guarantee the feasibility of all QP problems that need to be solved in deriving explicit solutions?

This paper resolves the QP feasibility problem above when an OCBF controller is used in decentralized merging control, thus ensuring that feasible trajectories are always possible. The merging control problem is formulated as in \cite{xiao2021decentralized} to jointly minimize the travel time and energy consumption subject to speed-dependent safety constraints as well as vehicle limitations. We adopt the OCBF approach and guarantee the feasibility of each QP problem by adding a single feasibility constraint to it following the strategy developed in \cite{Xiao2021} for general optimal control problems. While the feasibility constraints constructed in \cite{Xiao2021} are limited to be independent of the control, in this paper we exploit the structure of the safety constraints in the merging problem to derive control-dependent feasibility constraints and prove that all QPs needed to fully solve the merging problem are feasible. 

The paper is organized as follows. In Section II, the formulation of the merging control problem is reviewed. In Section III, we explain how to transition from an optimal control solution to the OCBF controller using a sequence of QPs. In Section IV, we derive the new constraints added to the QPs and prove that this ensures their feasibility. Simulation results are provided in Section V illustrating how to prevent CAV trajectories from becoming eventually infeasible by using the new feasibility constraints included in the OCBF controller.

\section{PROBLEM FORMULATION AND APPROACH}

\begin{figure}[ptbh]
\centering
\includegraphics[scale=0.22]{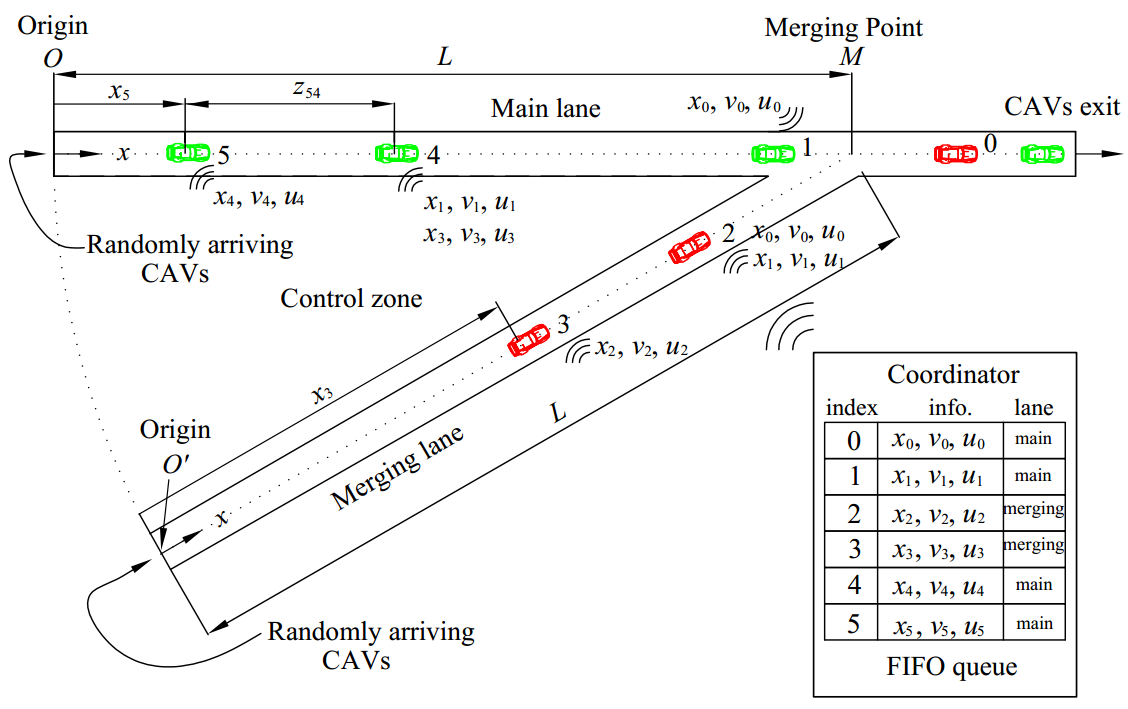} \caption{The merging problem.
{CAVs randomly arrive at the origins of the main and merging roads.
Collisons may occur at the MP. A coordinator is associated with the MP to
maintain the FIFO queue and share information among CAVs as needed.}}%
\label{modelF}%
\end{figure}

The merging problem arises when traffic must be joined from two different
roads, usually associated with a main lane and a merging lane as shown in
Fig.\ref{modelF}. We consider the case where all traffic consists of CAVs
randomly arriving at the two roads joined at the Merging Point (MP) $M$ where a collision may
occur. A coordinator is associated with the MP whose function is to maintain a
First-In-First-Out (FIFO) queue of CAVs based on their arrival time at the CZ
and enable real-time Vehicle-to Infrastructure (V2I) communication with the CAVs that are in the CZ, as well as the last one leaving the CZ. The segment from the origin $O$ or $O^{\prime}$ to the MP $M$ has a length $L$ for both roads, where $L$ is selected to be as large as possible
subject to the coordinator's communication range and the road network's
configuration and it defines the CZ. Since we consider
single-lane roads in this merging problem, CAVs may not overtake each other in the CZ (extensions to multi-lane merging are given in \cite{XiaoITSC2020}).  The FIFO assumption imposed so that CAVs cross the MP in their order of arrival is made for simplicity (and often to
ensure fairness), but can be relaxed through dynamic resequencing schemes as in \cite{XiaoACC2020} where optimal solutions are similarly derived but for different selected CAV sequences.

Let $S(t)$ be the set of FIFO-ordered indices of all CAVs located in the CZ at
time $t$ along with the CAV (whose index is 0 as shown in Fig.\ref{modelF})
that has just left the CZ. Let $N(t)$ be the cardinality of $S(t)$. Thus, if a
CAV arrives at time $t$ it is assigned the index $N(t)$. All CAV indices in
$S(t)$ decrease by one when a CAV passes over the MP and the vehicle whose
index is $-1$ is dropped.

The vehicle dynamics for each CAV $i\in S(t)$ along the lane to which it
belongs take the form
\begin{equation}
\left[
\begin{array}
[c]{c}%
\dot{x}_{i}(t)\\
\dot{v}_{i}(t)
\end{array}
\right]  =\left[
\begin{array}
[c]{c}%
v_{i}(t)\\
u_{i}(t)
\end{array}
\right]  \label{VehicleDynamics}%
\end{equation}
where $x_{i}(t)$ denotes the distance to the origin $O$ ($O^{\prime}$) along
the main (merging) lane if the vehicle $i$ is located in the main (merging)
lane, $v_{i}(t)$ denotes the velocity, and $u_{i}(t)$ denotes the control
input (acceleration). We consider two objectives for each CAV subject to three
constraints, as detailed next.

$\textbf{Objective 1}$ (Minimizing travel time): Let $t_{i}^{0}$ and
$t_{i}^{m}$ denote the time that CAV $i\in S(t)$ arrives at the origin $O$ or
$O^{\prime}$ and the MP $M$, respectively. We wish to minimize the travel time
$t_{i}^{m}-t_{i}^{0}$ for CAV $i$.

$\textbf{Objective 2}$ (Minimizing energy consumption): We also wish to
minimize energy consumption for each CAV $i\in S(t)$ expressed as
\begin{equation}
J_{i}(u_{i}(t))=\int_{t_{i}^{0}}^{t_{i}^{m}}C(u_{i}(t))dt,
\end{equation}
where $C(\cdot)$ is a strictly increasing function of its argument.

$\textbf{Constraint 1}$ (Safety constraints between $i$ and $i_{p}$): Let
$i_{p}$ denote the index of the CAV which physically immediately precedes $i$
in the CZ (if one is present). We require that the distance $z_{i,i_{p}%
}(t):=x_{i_{p}}(t)-x_{i}(t)$ be constrained so that
\begin{equation}
z_{i,i_{p}}(t)\geq\varphi v_{i}(t)+\delta,\text{ \ }\forall t\in\lbrack
t_{i}^{0},t_{i}^{m}], \label{Safety}%
\end{equation}
where $v_{i}(t)$ is the speed of CAV $i\in S(t)$ and $\varphi$ denotes the reaction time (as a rule, $\varphi=1.8$s is used, e.g., \cite{Vogel2003}). If we define $z_{i,i_{p}}$ to be the distance from
the center of CAV $i$ to the center of CAV $i_{p}$, then $\delta$ is a
constant determined by the length of these two CAVs (generally dependent on
$i$ and $i_{p}$ but taken to be a constant for simplicity).

$\textbf{Constraint 2}$ (Safe merging (terminal constraint) between $i$ and
$i-1$): When $i-1 = i_p$, this constraint is redundant since (\ref{Safety}) is enforced, but when $i-1 \ne i_p$ there should be enough safe space at the MP $M$ for a merging CAV to
cut in, i.e.,
\begin{equation}
z_{i,i-1}(t_{i}^{m})\geq\varphi v_{i}(t_{i}^{m
})+\delta. \label{SafeMerging}%
\end{equation}

$\textbf{Constraint 3}$ (Vehicle limitations): Finally, there are constraints
on the speed and acceleration for each $i\in S(t)$, i.e.,
\begin{align} 
v_{\min} \leq v_i(t)\leq v_{\max}, \forall t\in[t_i^0,t_i^m],\\
u_{i,\min}\leq u_i(t)\leq u_{i,\max}, \forall t\in[t_i^0,t_i^m],  \label{VehicleConstraints}%
\end{align}
where $v_{\max}>0$ and $v_{\min}\geq0$ denote the maximum and minimum speed
allowed in the CZ, while $u_{i,\min}<0$ and $u_{i,\max}>0$ denote the minimum
and maximum control input, respectively.

\textbf{Optimization Problem Formulation. }Our goal is to determine a control
law {(as well as optimal merging time $t_{i}^{m}$)} to achieve
objectives 1-2 subject to constraints 1-3 for each $i\in S(t)$ governed by the
dynamics (\ref{VehicleDynamics}). The common way to minimize energy
consumption is by minimizing the control input effort $u_{i}^{2}(t)$, noting that the OCBF method
allows for more elaborate fuel consumption models, e.g., as
in \cite{Kamal2013}.
By normalizing travel time and $u_{i}^{2}(t)$, and using $\alpha\in\lbrack0,1)$,
we construct a convex combination as follows: {
\begin{equation}\small
\setlength{\abovedisplayskip}{2pt}\setlength{\belowdisplayskip}{2pt}\begin{aligned}J_i(u_i(t),t_i^m)= \int_{t_i^0}^{t_i^m}\left(\alpha + \frac{(1-\alpha)\frac{1}{2}u_i^2(t)}{\frac{1}{2}\max \{u_{i,\max}^2, u_{i,\min}^2\}}\right)dt \end{aligned}.\label{eqn:energyobja}%
\end{equation}
}If $\alpha=1$, then we solve (\ref{eqn:energyobja}) as a minimum time
problem. Otherwise, by defining $\beta:=\frac{\alpha\max\{u_{i,\max}%
^{2},u_{i,\min}^{2}\}}{2(1-\alpha)}$ and multiplying (\ref{eqn:energyobja}) by
$\frac{\beta}{\alpha}$, we have: {\small
\begin{equation}
\setlength{\abovedisplayskip}{1pt}\setlength{\belowdisplayskip}{1pt}J_{i}%
(u_{i}(t),t_{i}^{m}):=\beta(t_{i}^{m}-t_{i}^{0})+\int_{t_{i}^{0}}^{t_{i}^{m}%
}\frac{1}{2}u_{i}^{2}(t)dt,\label{eqn:energyobj}%
\end{equation}
}
where $\beta\geq0$ is a weight factor that can be adjusted to penalize
travel time relative to the energy cost, {subject to
(\ref{VehicleDynamics}), (\ref{Safety})-(\ref{VehicleConstraints}) and the
terminal constraint $x_{i}(t_{i}^{m})=L$, given $t_{i}^{0},x_{i}(t_{i}%
^{0}),v_{i}(t_{i}^{0})$.}

\section{Optimal Control and Control Barrier Function Controller}
This merging control problem can be analytically solved, however, as pointed out in \cite{xiao2020bridging}, it will become computationally intensive when one of more constraints become active. To obtain a solution in real time while guaranteeing that no safety constraint is violated, the OCBF approach \cite{xiao2020bridging} is adopted through the following steps: (i) an optimal control solution for the \emph{unconstrained} optimal control problem is first obtained as a reference control, (ii) the resulting reference trajectory is optimally tracked subject to the control bounds \eqref{VehicleConstraints} as well as a set of CBF constraints enforcing \eqref{Safety}, \eqref{SafeMerging}. Using the forward invariance property of CBFs \cite{Xiao2019}, these constraints are guaranteed to be satisfied at all times if they are initially satisfied. The importance of CBFs is that they impose linear constraints on the control which, if satisfied, guarantee the satisfaction of the associated original constraints that involve the state and/or control. This comes at the expense of potential conservativeness in the control since the CBF constraint is a sufficient condition for ensuring its associated original problem constraint. This whole process is efficiently carried out in a decentralized way and leads to a sequence of QPs solved over discrete time steps, since the objective function is quadratic and the CBF constraints are linear in the control.

\textbf{Unconstrained optimal control solution}: With all constraints
inactive (including at $t_{i}^{0}$), the solution of \textbf{Problem 1} is achieved by standard Hamiltonian analysis \cite{Bryson1969} so that, as shown in \cite{xiao2021decentralized}, the optimal control, velocity
and position trajectories of CAV $i$ have the form: 
\begin{align}
u_{i}^{\ast }(t)& =a_{i}t+b_{i}  \label{equ:u} \\
v_{i}^{\ast }(t)& =\frac{1}{2}a_{i}t^{2}+b_{i}t+c_{i}  \label{equ:v} \\
x_{i}^{\ast }(t)& =\frac{1}{6}a_{i}t^{3}+\frac{1}{2}b_{i}t^{2}+c_{i}t+d_{i}
\label{equ:x}
\end{align}%
where the parameters $a_{i}$, $b_{i}$, $c_{i}$, $d_{i}$ and $t_{i}^{m}$ are
obtained by solving a set of nonlinear algebraic equations: 
\begin{equation}
\begin{split}
& \frac{1}{2}a_{i}\cdot (t_{i}^{0})^{2}+b_{i}\cdot t_{i}^{0}+c_{i}=v_{i}^{0},
\\
& \frac{1}{6}a_{i}\cdot (t_{i}^{0})^{3}+\frac{1}{2}b_{i}\cdot
(t_{i}^{0})^{2}+c_{i}t_{i}^{0}+d_{i}=0, \\
& \frac{1}{6}a_{i}\cdot (t_{i}^{m})^{3}+\frac{1}{2}b_{i}\cdot
(t_{i}^{m})^{2}+c_{i}t_{i}^{m}+d_{i}=x_{i}(t_{i}^{m}), \\
& a_{i}t_{i}^{m}+b_{i}=0, \\
& \beta +\frac{1}{2}a_{i}^{2}\cdot
(t_{i}^{m})^{2}+a_{i}b_{i}t_{i}^{m}+a_{i}c_{i}=0.
\end{split}%
\end{equation}

This set of equations only needs to be solved when CAV $i$ enters the CZ
and, as already mentioned, it can be done very efficiently.

\textbf{Optimal tracking controller with CBFs}: Once we obtain the
unconstrained optimal control solutions \eqref{equ:u}-\eqref{equ:x}, we use
a function $h(u_{i}^{\ast }(t),x_{i}^{\ast }(t),x_{i}(t))$ as a control
reference $u_{ref}(t)=h(u_{i}^{\ast }(t),x_{i}^{\ast }(t),x_{i}(t))$, where 
$x_{i}(t)$ provides feedback from the actual observed CAV trajectory. We
then design a controller that minimizes $\int_{t_{i}^{0}}^{t_{i}^{m}}\frac{1%
}{2}(u_{i}(t)-u_{\mathrm{ref}}(t))^2dt$ subject to all constraints %
\eqref{Safety}, \eqref{SafeMerging} and \eqref{VehicleConstraints}%
. This is accomplished as reviewed next (see also \cite{xiao2020bridging}).

First, let $\bm x_{i}(t)\equiv (x_{i}(t),v_{i}(t))$. Due to the vehicle
dynamics \eqref{VehicleDynamics}, define $f(\bm x_{i}(t))=[v_{i}(t),0]^{T}$ and 
$g(\bm x_{i}(t))=[0,1]^{T}$. The constraints \eqref{Safety} and
\eqref{SafeMerging} are expressed in the
form $b_{k}(\bm x_{i}(t))\geq 0,k\in \{1,...,B\}$ where $B$ is the number of
constraints. The CBF method maps the constraint $b_{k}(\bm x_{i}(t))\geq 0$
to a new constraint which directly involves the control $u_{i}(t)$ and takes
the (linear in the control) form 
\begin{equation}
L_{f}b_{k}(\bm x_{i}(t))+L_{g}b_{k}(\bm x_{i}(t))u_{i}(t)+\gamma (b_{k}(\bm %
x_{i}(t)))\geq 0,  \label{equ:cbf}
\end{equation}%
where $L_{f},L_{g}$ denote the Lie derivatives of $b_{k}(\bm x_{i}(t))$
along $f$ and $g$ respectively and $\gamma (\cdot )$ denotes some class-$%
\mathcal{K}$ function \cite{Xiao2019}. The forward invariance
property of CBFs guarantees that a control input that keeps \eqref{equ:cbf}
satisfied will also enforce $b_{k}(\bm x_{i}(t))\geq 0$. In other words, the
constraints \eqref{Safety}, \eqref{SafeMerging} are never violated.

To optimally track the reference speed trajectory, a CLF function $V(\bm %
x_{i}(t))$ is used. Letting $V(\bm x_{i}(t))=(v_{i}(t)-v_{\mathrm{ref}}(t))^{2}$, the
CLF constraint takes the form 
\begin{equation}
L_{f}V(\bm x_{i}(t))+L_{g}V(\bm x_{i}(t))u_{i}(t)+\epsilon V(x_{i}(t))\leq
e_{i}(t),  \label{equ:clf}
\end{equation}%
where $\epsilon >0$, and $e_{i}(t)$ is a relaxation variable which makes the
constraint soft. Then, the OCBF controller optimally tracks the reference
trajectory by solving the optimization problem: 
\begin{equation}
\min_{u_{i}(t),e_{i}(t)}\int_{t_{i}^{0}}^{t_{i}^{m}}\left( \beta
e_{i}^{2}(t)+\frac{1}{2}(u_{i}(t)-u_{\mathrm{ref}(t)})^2\right)   \label{equ:ocbf}
\end{equation}%
subject to the CBF constraints %
\eqref{equ:cbf}, the CLF constraints \eqref{equ:clf} and the control bounds \eqref{VehicleConstraints}. There are several
possible choices for $u_{\mathrm{ref}}(t)$ and $v_{\mathrm{ref}}(t)$. In the sequel, we choose
the following referenced trajectory with feedback included to reduce the tracking position error: 
\begin{equation}
v_{\mathrm{ref}}(t)=\frac{x_i^{\ast}(t)}{x_i(t)}v_{i}^{\ast }(t),\text{ \ \ }u_{\mathrm{ref}}(t)=\frac{x_i^{\ast}(t)}{x_i(t)}u_{i}^{\ast }(t)
\end{equation}%
where $u_{i}^{\ast}(t)$, $v_{i}^{\ast }(t)$ and $x_{i}^{\ast }(t)$ are obtained from %
\eqref{equ:u}, \eqref{equ:v} and \eqref{equ:x}.

We can solve problem \eqref{equ:ocbf} by discretizing $[t_{i}^{0},t_{i}^{m}]$ into
intervals of equal length $\Delta t$ and solving \eqref{equ:ocbf} over each
time interval. The decision variables $u_{i}(t)$ and $e_{i}(t)$ are assumed
to be constant on each such time interval $[t_{i}^{0} + k\Delta t, t_{i}^{0} + (k+1)\Delta t]$ and can be easily obtained by
solving a Quadratic Program (QP) problem \eqref{equ:ocbf_qp} since all CBF constraints are
linear in the decision variables $u_{i}(t)$ and $e_{i}(t)$ (fixed over each interval $[t_{i}^{k}, t_{i}^{k} + \Delta t]$). 
\begin{equation}
\label{equ:ocbf_qp}
\begin{split}
\min_{u_{i}(t) ,e_{i}(t)}&~\beta e_{i}(t)^{2}+\frac{1}{2}(u_{i}(t)-u_{\mathrm{ref}}(t))^2 \\
s.t.~&~ \eqref{equ:cbf}, \eqref{equ:clf}, \eqref{VehicleConstraints} \\
&~ t = t_i^0 + k\Delta t
\end{split}
\end{equation}%

By repeating this process until CAV $i$ exits the CZ, the solution to \eqref{equ:ocbf} is
obtained \emph{if \eqref{equ:ocbf_qp} is feasible for each time interval}. This approach is simple and computationally efficient. However, it is also myopic since each QP is solved over a single time step,
which often leads to infeasible QPs at future time steps, especially when the control bound \eqref{VehicleConstraints} is tight.

\section{Feasibility Guaranteed OCBF}
To avoid the infeasibility caused by the myopic QP solving approach in the CBF method, an additional ``feasibility constraint'' $b_F(\bm x_i(t), u_i(t)) \geq 0$ is introduced in \cite{Xiao2021}. A feasibility constraint is defined as a constraint that makes the QP corresponding to the next time interval feasible, thus, in the case of \eqref{equ:ocbf_qp}, a feasibility constraint has the following properties: (i) it guarantees that \eqref{equ:cbf} and \eqref{VehicleConstraints} do not conflict, (ii) the feasibility constraint itself conflicts with neither \eqref{equ:cbf} nor \eqref{VehicleConstraints}. In general, we call any two state and/or control constraints (e.g., any CBF constraint) \emph{conflict-free} if their intersection is non-empty in terms of the control.

In \cite{Xiao2021}, a general sufficient condition for feasibility is provided based on the assumption that the feasibility constraint $b_F(\bm x(t), u(t))$ is independent of the control $u(t)$. This assumption is hard to meet when it comes to the merging control problem, thus making such a sufficient condition difficult to determine. In what follows, we show that it is possible to find a feasibility constraint $b_F(\bm x_i(t), u(t)) \geq 0$ for each CAV $i$ without this assumption and explicitly derive this constraint which can provably guarantee feasibility. The detailed analysis and proofs are provided in the following subsections.

In the merging control problem, the form of the safety constraint depends on whether CAV $i$ and CAV $i - 1$ are in the same road. If CAV $i$ and $i-1$ are in the same road (i.e., $i-1=i_p$), the rear-end safety constraint needs to be considered. If CAV $i$ and $i-1$ are in different roads, the safe merging constraint (\ref{SafeMerging}) must be included. We will discuss the feasibility constraints corresponding to the two cases separately in what follows.

\subsection{Rear-end Safety Constraint}
\label{sec:re_safety}
When CAV $i$ and $i-1=i_p$ are in the same road, only the rear-end safety constraint needs to be considered: 
\begin{equation}
\label{equ:CBF_constraint1}
    b_1(\bm x_i(t)) = z_{i, i_p}(t) - \varphi v_i(t) + \delta \geq 0
\end{equation}

As $b_1(\bm x_i(t))$ is  differentiable, we can calculate the Lie derivatives $L_f(b_1(\bm x_i(t))) = v_{i_p} - v_i(t)$, $L_g(b_1(\bm x_i(t))) = -\varphi$. Applying \eqref{equ:cbf} and choosing a linear function $\gamma(x) = k_1 x$ as the class $\mathcal{K}$ function, the rear-end safety constraint \eqref{equ:CBF_constraint1} can be directly transformed into the following CBF constraint. 
\begin{equation}
    b_{\mathrm{cbf}_1}(\bm x_i, u_i) = v_{i_p} - v_i - \varphi u_i + k_1b_1(\bm x_i) \geq 0 \label{equ:cbf1}
\end{equation}


As $-L_g(b_1(x_i(t))) = \varphi > 0$, multiplying both sides of the control bound \eqref{VehicleConstraints} by $-L_g(b_1(x_i(t)))$ will not change the direction of the inequalities. Hence, we have
\begin{equation}
\label{equ:acc_constraint1}
    \varphi u_{i, \min} \leq \varphi u_i(t) \leq \varphi u_{i, \max}
\end{equation}

Note that \eqref{equ:cbf1} can be rewritten as 
\begin{equation}
\label{equ:cbf1_trans}
    \varphi u_i(t) \leq v_{i_p}(t) - v_i(t) + k_1b_1(\bm x_i(t))
\end{equation}
where $\varphi u_i(t) \leq u_{i, \max}$ never conflicts with \eqref{equ:cbf1_trans} as they have the same inequality direction. Thus, we can guarantee that \eqref{equ:cbf1} and \eqref{equ:acc_constraint1} are conflict-free by adding a feasibility constraint
\begin{equation}
\label{equ:fc_1}
   b_F(\bm x_i(t)) = v_{i_p}(t) - v_i(t) + k_1b_1(\bm x_i(t)) - \varphi u_{i, \min} \geq 0
\end{equation}
We can now consider this feasibility constraint as a new CBF and apply \eqref{equ:cbf} to transform it into a CBF constraint. Choosing a linear function $\gamma(x) = k_F x$ as the class $\mathcal{K}$ function, the corresponding CBF constraint is 
\begin{equation}
\label{equ:fc1}
\begin{split}
    &u_{i_p} - u_i + k_1(v_{i_p} - v_i - \varphi u_i) + k_F(v_{i_p} - v_i)\\
    & + k_F k_1(z_{i, i_p} - \varphi v_i + \delta) - k_F\varphi u_{i, \min} \geq 0
\end{split}
\end{equation}
where the argument $t$ of the functions above is omitted for simplicity.

Next, we determine a feasible constraint to be added to every QP so that it guarantees the QP of the next time interval is feasible. This is done by choosing $k_F$ so that $k_F = k_1$ and \eqref{equ:fc1} becomes
\begin{equation}
\label{equ:fc1_trans}
\begin{split}
    &u_{i_p} - u_i + k_1(v_{i_p} - v_i - \varphi u_{i, \min}) \\
    &+ k_1(v_{i_p} - v_i - \varphi u_i + k_1(z_{i, i_p} - \varphi v_i + \delta)) \geq 0
\end{split}
\end{equation}

Define a ``candidate function'' $\eta(\bm x_i, u_i)$ \cite{Xiao2021} as
\begin{equation}
\label{equ:candidate}
    \eta_1(\bm x_i, u_i) = u_{i_p} - u_i + k_1(v_{i_p} - v_i - \varphi u_{i, \min})
\end{equation}
Then, replacing the first three terms of the feasibility CBF constraint \eqref{equ:fc1_trans} with $\eta_1(\bm x_i, u_i)$ and noting that the remaining terms are given by $b_{\mathrm{cbf}_1}(\bm x_i, u_i)$ defined in \eqref{equ:cbf1}, to substitute the second row with $b_{\mathrm{cbf}_1}(\bm x_i, u_i)$, \eqref{equ:fc1_trans} becomes 
\begin{equation}
\label{equ:fc1_cbf2}
    \eta_1(\bm x_i, u_i) + k_1b_{\mathrm{cbf}_1}(\bm x_i, u_i) \geq 0
\end{equation}
Since $b_{\mathrm{cbf}_1}(\bm x_i, u_i) \geq 0$ is required in \eqref{equ:cbf1}, it follows that \eqref{equ:fc1_cbf2} will be satisfied if 
\begin{equation}
\label{equ:sc_rear}
    \eta_1(\bm x_i, u_i) \geq 0
\end{equation}
Setting
\begin{equation}
\label{equ:betaeta}
    b_{\eta_1}(\bm x_i) = v_{i_p} - v_i - \varphi u_{i, \min}
\end{equation}
in \eqref{equ:candidate}, we can view $b_{\eta_1}(\bm x_i)$ as a CBF and apply \eqref{equ:cbf} to observe that the corresponding CBF constraint coincides with \eqref{equ:sc_rear}. Adding the CBF constraint \eqref{equ:sc_rear} to the QP \eqref{equ:ocbf_qp}, we will show next that \eqref{equ:sc_rear} is a constraint that guarantees the feasibility of the QP corresponding to the next time interval. Before establishing this result, we make the following two assumptions.

\begin{assumption}
\label{assume:u_min}
    All CAVs have the same minimum acceleration, i.e. $u_{i, \min} = u_{\min}$.
\end{assumption}

This assumption is required to guarantee that \eqref{equ:sc_rear} and \eqref{VehicleConstraints} are conflict-free. It a weak assumption and can be easily enforced since all CAVs are operating within the same CZ, i.e., they can reach agreement on a common $u_{\min}=\min_i \{u_{i, \min}\}$ through V2V communication.


\begin{assumption}
\label{assume:small_t}
    $\Delta t$ is adequately small such that the forward invariance property of CBFs remains in force.
\end{assumption}

This assumption is made to utilize the forward invariance property of CBFs to guarantee safety. It can be met by decreasing the time interval or by using the recently proposed  event-driven technique \cite{Xiao2021D} which uses a tunable inter-QP interval (instead of a fixed time-driven one) which is guaranteed to preserve constraint satisfaction.


\begin{theorem}
\label{theorem:rear1}
    If $b_{\eta_1}(\bm x(t)) \geq 0$ and the QP \eqref{equ:ocbf_qp} subject to \eqref{equ:cbf1}, \eqref{VehicleConstraints} and \eqref{equ:sc_rear} is feasible at time $t$, then the QP corresponding to time $t + \Delta t$ is also feasible.
\end{theorem}

\textbf{Proof:} By \emph{Assumption \ref{assume:u_min}}, $\min_i\{u_{i}\} = u_{\min} \leq u_{i_p}(t)$, thus there always exists a control input $u_i(t) \in [u_{\min}, u_{i, \max}]$ such that $u_{i_p}(t) - u_i(t) \geq 0$. As $b_{\eta_1}(\bm x(t)) \geq 0$, applying \eqref{equ:candidate}, we can always find a feasible control $u_i(t)$ such that $\eta_1(\bm x_i(t), u_i(t)) \geq 0$. As $\eta_1(\bm x_i(t), u_i(t)) \geq 0$ is the CBF constraint corresponding to $b_{\eta_1}(\bm x_i(t)) \geq 0$, using the forward invariance property of CBFs under \emph{Assumption \ref{assume:small_t}}, we have $b_{\eta_1}(\bm x(t + \Delta t)) \geq 0$. Thus, there always exists a control $u_i(t)\in [u_{\min}, u_{i, \max}]$ such that $\eta_1(\bm x_i(t +\Delta t), u_i(t + \Delta t)) \geq 0$. Hence, \eqref{equ:sc_rear} and \eqref{VehicleConstraints} are conflict-free at $t +\Delta t$.


Since $\varphi \geq 0$, \eqref{equ:cbf1} constrains the control $u_i(t + \Delta t)$ with an upper bound. Similarly, $u_i(t + \Delta t)$ is also constrained by an upper bound through \eqref{equ:sc_rear}. Thus, \eqref{equ:cbf1} and \eqref{equ:sc_rear} are conflict-free at $t +\Delta t$.

Since the QP \eqref{equ:ocbf_qp} subject to \eqref{equ:cbf1}, \eqref{VehicleConstraints} and \eqref{equ:sc_rear} is feasible at time $t$, it follows that $b_{\mathrm{cbf}_1}(\bm x_i(t), u_i(t)) \geq 0$, $b_F(\bm x(t)) \geq 0$. As $\eta_1(\bm x_i(t), u_i(t)) \geq 0$ always has a solution $u_i(t) \in [u_{\min}, u_{i, \max}]$, there exists a control under which \eqref{equ:fc1_cbf2} is satisfied. Since \eqref{equ:fc1_cbf2} is the CBF constraint of \eqref{equ:fc_1},  using the forward invariance of CBFs under \emph{Assumption \ref{assume:small_t}}, we have $b_F(\bm x(t + \Delta t)) \geq 0$, which implies that \eqref{equ:cbf1} and \eqref{VehicleConstraints} are conflict-free at time $t + \Delta t$.

Thus, all constraints of the QP \eqref{equ:ocbf_qp} are conflict-free at $t + \Delta t$ and the QP corresponding to time $t+\Delta$ is feasible. \hfill$\blacksquare$

\begin{assumption}
\label{assume:initial}
    The following initial conditions are satisfied: $b_1(\bm x_i(t_i^0)) \geq 0, b_F(\bm x_i(t_i^0)) \geq 0, b_{\eta_1}(\bm x_i(t_i^0)) \geq 0$
\end{assumption}

The constraint $b_1(\bm x_i(t_i^0)) \geq 0$ requires CAV $i$ to meet the rear-end safety with the immediately preceding CAV (if one exists) when entering the CZ. In addition, $b_F(\bm x_i(t_i^0)) \geq 0$ requires that the CBF constraint is initially conflict-free with the control bounds and $b_{\eta_1}(\bm x_i(t_i^0)) \geq 0$ indicates that CAV $i$ should not be too faster than the preceding CAV. These constraints are reasonable and can be met using a Feasibility Enforcement Zone (FEZ) \cite{zhangACC2017} that precedes the CZ. 

\begin{theorem}
\label{theorem:rear2}
    Under \emph{Assumptions 1,2,3}, the QP \eqref{equ:ocbf_qp} subject to \eqref{equ:cbf1}, \eqref{VehicleConstraints} and \eqref{equ:sc_rear} corresponding to any time interval $[t_{i}^{0} + k\Delta t, t_{i}^{0} + (k+1)\Delta t] \subset [t_i^0, t_i^m]$ is feasible.
\end{theorem}

\textbf{Proof:} Since $b_F(\bm x_i(t_i^0)) \geq 0$, the CBF constraint \eqref{equ:cbf1} does not conflict with the control bounds \eqref{VehicleConstraints}. Using \eqref{equ:candidate} under $b_{\eta_1}(\bm x_i(t_i^0)) \geq 0$ and \emph{Assumption \ref{assume:u_min}}, there always exists a control $u_i(t_i^0) \in [u_{\min}, u_{i,\max}]$ such that \eqref{equ:sc_rear} is satisfied, which indicates that \eqref{equ:sc_rear} and \eqref{VehicleConstraints} are conflict-free. As \eqref{equ:cbf1} and \eqref{equ:sc_rear} have the same inequality direction for $u_i(t_i^0)$, they are conflict-free as well. Thus, the QP \eqref{equ:ocbf_qp} subject to \eqref{equ:cbf1}, \eqref{VehicleConstraints} and \eqref{equ:sc_rear} is feasible at time $t_i^0$. 

According to \emph{Theorem} \ref{theorem:rear1}, the QP corresponding to time $t_i^0 + \Delta t$ is feasible. \emph{Assumption \ref{assume:small_t}} preserves the forward invariance property of CBFs which guarantees that $b_1(\bm x_i(t_i^0 + \Delta t)) \geq 0$ and $b_{\eta_1}(\bm x(t_i^0 + \Delta t))\geq 0$ under \emph{Assumption \ref{assume:initial}}.

Following the same procedure, it is easy to prove that if the QP is feasible at $t_i^0 + k\Delta t$, $b_1(\bm x_i(t_i^0 + k\Delta t)) \geq 0$ and  $b_{\eta_1}(\bm x_i(t_i^0 + k\Delta t)) \geq 0$, then the QP corresponding to $t_i^0 + (k + 1)\Delta t$ is feasible, $b_1(\bm x_i(t_i^0 + (k + 1)\Delta t)) \geq 0$ and $b_{\eta_1}(\bm x_i(t_i^0 + (k + 1)\Delta t)) \geq 0$ and the proof of the theorem is completed by induction. \hfill$\blacksquare$

\subsection{Safe Merging Constraint}
When CAVs $i$ and $i - 1$ are in different roads, they should also satisfy the merging safety constraint
\begin{equation}
\label{equ:merging_constraint}
    z_{i, i-1}(t_i^m) - \varphi v_i(t_i^m) - \delta \geq 0
\end{equation}
This constraint differs from the rear-end safety constraint in that it only applies to specific time instants $t_{i}^{m}$. This poses a technical complication due to the fact that a CBF must always be in a continuously differentiable form. We can convert \eqref{equ:merging_constraint} to such a form using a technique similar to the one used in \cite{xiao2020bridging} to define 
\begin{equation}
    b_2(\bm x_i(t)) = z_{i, i-1}(t) - \Phi(\bm x_i(t))v_i(t) - \delta \geq 0
\end{equation}
where $\Phi(\bm x_i(t)) = \frac{\varphi}{x_i(t_i^m)}x_i(t)$. Note that $\Phi(\bm x_i(t_i^m)) = \varphi$ consistent with (\ref{equ:merging_constraint}).

Setting $\varphi_2 = \frac{\varphi}{x_i(t_i^m)}$ and omitting the time argument $t$ to ease notation, the Lie derivatives for $b_2(\bm x_i(t))$ are $L_f(b_2(\bm x_i)) = v_{i-1} - v_i - \varphi_2 v_i^2$ and $L_g(b_2(\bm x_i)) = -\varphi_2 x_i$. Thus, using \eqref{equ:cbf} and choosing $\gamma(x) = k_2 x$, \eqref{equ:merging_constraint} is mapped onto the CBF constraint
\begin{equation}
\small
\label{equ:cbf_merging}
    b_{\mathrm{cbf}_2}(\bm x_i, u_i) = v_{i - 1} - v_i - \varphi_2 v_i^2 - \varphi_2 x_iu_i + k_2b_2(\bm x_i) \geq 0
\end{equation}
As $-L_g(b_2(\bm x_i)) = \varphi_2 x_i \geq 0$, multiplying both sides of the control bound \eqref{VehicleConstraints} by $-L_g(b_2(\bm x_i))$ will not change the direction of the inequalities, thus, under \emph{Assumption \ref{assume:u_min}},
\begin{equation}
    \varphi_2 x_i u_{\min} \leq \varphi_2 x_i u_i\leq \varphi_2 x_i u_{i, \max}
\end{equation}

Similar to Sec. \ref{sec:re_safety}, since \eqref{equ:cbf_merging} requires $\varphi_2 x_iu_i$ to be smaller than an upper bound, \eqref{equ:cbf_merging} can only possibly conflict with the lower bound $\varphi_2 x_i u_{\min}$. This possible conflict can be avoided by adding a feasibility constraint 
\begin{equation}
\label{equ:fc2}
    b_F(\bm x_i) = v_{i - 1} - v_i - \varphi_2 v_i^2 + k_2b_2(\bm x_i) - \varphi_2 x_i u_{\min} \geq 0
\end{equation}

Choosing a linear function $\gamma(x) = k_F x$ as the class $\mathcal{K}$ function and applying \eqref{equ:cbf}, the feasibility constraint \eqref{equ:fc2} is transformed into a CBF constraint
\begin{equation}
\begin{split}
\label{equ:fc2_cbf}
    &k_2(v_{i - 1} - v_i - \varphi_2 v_i^2) - \varphi_2 v_i u_{\min} \\
    +& u_{i - 1} - u_i - 2\varphi_2 v_i u_i + k_2\varphi_2 x_i u_i \\
    +& k_F(v_{i - 1} - v_i - \varphi_2 v_i^2 - \varphi_2 x_i u_i) + k_F k_2 b_2(\bm x_i) \\
    -& k_F\varphi_2 x_i u_{\min} \geq 0
\end{split}
\end{equation}
Selecting $k_F = k_2$, we are able to determine the constraint that guarantees the feasibility of the QP at the next time interval. In this case, \eqref{equ:fc2_cbf} becomes
\begin{equation}
\begin{split}
    \label{equ:fc2_cbf_trans}
    &u_{i - 1} - u_i - 2\varphi_2 v_i u_i - \varphi_2 v_i u_{\min} \\
     +& k_2(v_{i - 1} - v_i - \varphi_2 v_i^2 - \varphi_2 x_i u_{\min}) \\ 
     +& k_2(v_{i - 1} - v_i - \varphi_2 v_i^2 - \varphi_2 x_i u_i + k_2b_2(\bm x_i)) \geq 0
\end{split}
\end{equation}
Define the first five terms of \eqref{equ:fc2_cbf_trans} as the ``candidate function'' 
\begin{equation}
\begin{split}
\label{equ:candidate2}
    \eta_2(\bm x_i, u_i) = &u_{i - 1} - u_i - 2\varphi_2 v_iu_i - \varphi_2 v_i u_{\min} \\
    &+ k_2(v_{i - 1} - v_i - \varphi_2 v_i^2 - \varphi_2 x_i u_{\min}).
\end{split}
\end{equation}
Then, using \eqref{equ:cbf_merging}, the constraint \eqref{equ:fc2_cbf_trans} can be rewritten as
\begin{equation}
    \eta_2(\bm x_i, u_i) + k_2(b_{\mathrm{cbf}_2}(\bm x_i, u_i)) \geq 0
\end{equation}
Setting
\begin{equation}
    b_{\eta_2}(\bm x_i) = v_{i - 1} - v_i - \varphi_2 v_i^2 - \varphi_2 x_i u_{\min}
\end{equation}
we can now view $b_{\eta}$ as a CBF and, applying \eqref{equ:cbf} with $\gamma(x) = k_2(x)$, the corresponding CBF constraint is 
\begin{equation}
\label{equ:sc_merging}
    \eta_2(\bm x_i, u_i) \geq 0
\end{equation}
Adding the CBF constraint \eqref{equ:sc_merging} to the QP \eqref{equ:ocbf_qp}, we show next that \eqref{equ:sc_merging} is a sufficient condition for the feasibility of \eqref{equ:ocbf_qp}.




\begin{lemma}
\label{lemma:1}
    If $v_i \geq 0, u_{\min} \leq 0$, there exists a control input $u_i(t) \in [u_{\min}, u_{i, \max}]$, such that
    \begin{equation}
        \label{equ:lemma1}
        u_{i - 1} - u_i - 2\varphi_2 v_i u_i - \varphi_2 v_i u_{\min} \geq 0
    \end{equation}
\end{lemma}

\textbf{Proof:}
Note that \eqref{equ:lemma1} can be rewritten as
\begin{equation}
\label{equ:lemma2}
    u_i + \frac{1}{2}u_{\min} \leq \frac{1}{1 + 2\varphi_2 v_i} (u_{i-1} + \frac{1}{2}u_{\min})
\end{equation}

Under \emph{Assumption \ref{assume:u_min}}, 
\begin{equation}
\label{equ:lemma1_ui-1}
    \frac{1}{1 + 2\varphi_2 v_i} (u_{i-1} + \frac{1}{2}u_{\min}) \geq \frac{1}{1 + 2\varphi_2 v_i}\frac{3}{2}u_{\min}
\end{equation}

As $v_i \geq 0$, we have $1 + 2\varphi_2 v_i \geq 1$. Additionally, $u_{\min} \leq 0$, thus
\begin{equation}
    u_{\min} \leq \frac{1}{1 + 2\varphi_2 v_i}u_{\min}
\end{equation}

As $\min_{u_i} (u_i + \frac{1}{2}u_{\min}) = \frac{3}{2} u_{\min}$, we have
\begin{equation}
\label{equ:lemma1_ui}
    \min_{u_i} (u_i + \frac{1}{2}u_{\min}) \leq \frac{1}{1 + 2\varphi_2 v_i}\frac{3}{2}u_{\min}
\end{equation}
Combining \eqref{equ:lemma1_ui} with \eqref{equ:lemma1_ui-1}, there always exists a control input $u_i(t)\in [u_{\min}, u_{i, \max}]$ such that \eqref{equ:lemma2} is satisfied. Hence, \emph{Lemma \ref{lemma:1}} is proved. \hfill $\blacksquare$


\begin{theorem}
\label{theorem:merging1}
    If $b_{\eta_2}(\bm x(t)) \geq 0$, $v_i \geq 0, u_{\min} \leq 0$ and the QP \eqref{equ:ocbf_qp} subject to \eqref{equ:cbf_merging}, \eqref{VehicleConstraints} and \eqref{equ:sc_merging} is feasible at time $t$, then the QP corresponding to time $t + \Delta t$ is also feasible.
\end{theorem}

\textbf{Proof:} According to \emph{Lemma} \ref{lemma:1}, there always exists a control input $u_i(t)\in[u_{\min}, u_{i, \max}]$ such that \eqref{equ:lemma1} is satisfied. As $b_{\eta_2}(\bm x_i(t)) \geq 0$, according to \eqref{equ:candidate2}, there exists a control input $u_i(t)\in[u_{\min}, u_{i, \max}]$ such that $\eta_2(\bm x_i(t), u_i(t)) \geq 0$. \emph{Assumption \ref{assume:small_t}} allows us to use the forward invariance property of CBFs to get $b_{\eta_2}(\bm x_i(t + \Delta t)) \geq 0$. Thus, there always exists a control input $u_i(t)\in[u_{\min}, u_{i, \max}]$ such that $\eta_2(\bm x_i(t +\Delta t), u_i(t +\Delta t)) \geq 0$. Hence, \eqref{equ:sc_merging} and \eqref{VehicleConstraints} are conflict-free at $t +\Delta t$.


As $\varphi_2 x_i \geq 0$, \eqref{equ:cbf_merging} constrains $u_i(t +\Delta t)$ with an upper bound. Similarly, $2\varphi_2 v_i \geq 0$, $u_i(t +\Delta t)$ is also constrained with an upper bound in \eqref{equ:sc_merging}. Thus, \eqref{equ:cbf_merging} and \eqref{equ:sc_merging} are conflict-free at $t +\Delta t$.

The QP \eqref{equ:ocbf_qp} subject to \eqref{equ:cbf_merging}, \eqref{VehicleConstraints} and \eqref{equ:sc_merging} is feasible at time $t$, thus $b_{\mathrm{cbf}_2}(\bm x_i(t), u_i(t)) \geq 0$, $b_F(\bm x(t)) \geq 0$. As $\eta_2(\bm x_i(t), u_i(t)) \geq 0$ always has a solution $u_i(t) \in [u_{\min}, u_{i, \max}]$, there exists a control under which \eqref{equ:fc2_cbf_trans} is satisfied. As \eqref{equ:fc2_cbf_trans} is the CBF constraint of \eqref{equ:fc2}, it follows from the forward invariance property of CBFs that $b_F(\bm x(t + \Delta t)) \geq 0$. This indicates that \eqref{equ:cbf_merging} and \eqref{VehicleConstraints} are conflict-free at time $t + \Delta t$.

Thus, all constraints of the QP \eqref{equ:ocbf_qp} are conflict-free at $t + \Delta t$ and the QP corresponding to time $t+\Delta$ is feasible. \hfill$\blacksquare$

\begin{assumption}
\label{assume:initial2}
    The following initial conditions are satisfied: $b_2(\bm x_i(t_i^0)) \geq 0, b_F(\bm x_i(t_i^0)) \geq 0, b_{\eta_2}(\bm x_i(t_i^0)) \geq 0$
\end{assumption}

This assumption is similar to \emph{Assumption \ref{assume:initial}}. The constraint $b_{\eta_2}(\bm x_i(t_i^0)) \geq 0$ requires CAV $i$ to be sufficiently slower than CAV $i-1$ upon entering the CZ. This restriction can still be met using a Feasibility Enforcement Zone (FEZ). 

\begin{theorem}
\label{theorem:merging2}
    Under \emph{Assumptions 1,2,4}, if $v_i \geq 0, u_{\min} \leq 0$, the QP \eqref{equ:ocbf_qp} subject to \eqref{equ:cbf_merging}, \eqref{VehicleConstraints} and \eqref{equ:sc_merging} corresponding to any time interval $[t_{i}^{0} + k\Delta t, t_{i}^{0} + (k+1)\Delta t] \subset [t_i^0, t_i^m]$ is feasible.
\end{theorem}

\textbf{Proof:} The proof is by induction following the same steps as in the proof of \emph{Theorem} \ref{theorem:rear2} and invoking \emph{Theorem \ref{theorem:merging1}}. \hfill$\blacksquare$


Note that when both the rear-end safety constraint and the safe merging constraint are included, the feasibility of the QP can still be guaranteed by adding one feasibility constraint corresponding to each CBF constraint, i.e. \eqref{equ:sc_rear} and \eqref{equ:sc_merging}.

\begin{theorem}
\label{theorem:rear_merging1}
    If $b_{\eta_1}(\bm x(t)) \geq 0$, $b_{\eta_2}(\bm x(t)) \geq 0$, $v_i \geq 0, u_{\min} \leq 0$, the QP \eqref{equ:ocbf_qp} subject to \eqref{equ:cbf1}, \eqref{equ:cbf_merging}, \eqref{VehicleConstraints}, \eqref{equ:sc_rear} and \eqref{equ:sc_merging} is feasible at time $t$, then the QP corresponding to time $t + \Delta t$ is also feasible.
\end{theorem}

\textbf{Proof:} \emph{Theorem \ref{theorem:rear1}} shows that \eqref{equ:cbf1}, \eqref{VehicleConstraints} and \eqref{equ:sc_rear} are conflict-free at $t+\Delta t$. \emph{Theorem \ref{theorem:merging1}} shows that \eqref{equ:cbf_merging}, \eqref{VehicleConstraints} and \eqref{equ:sc_merging} are conflict-free at $t+\Delta t$. In order to guarantee the feasibility of the QP corresponding to time $t+\Delta t$, we need to further prove \eqref{equ:cbf1}, \eqref{equ:cbf_merging}, \eqref{equ:sc_rear} and \eqref{equ:sc_merging} are conflict-free. By analyzing the coefficient of $u_i(t)$, we find $\varphi \geq 0$ in \eqref{equ:cbf1}, $1 \geq 0$ in \eqref{equ:sc_rear}, $\varphi_2 x_i \geq 0$ in \eqref{equ:cbf_merging} and $1+2\varphi_2 v_i \geq 0$ in \eqref{equ:sc_merging}. This indicates that all these constraints have the same inequality direction, thus \eqref{equ:cbf1}, \eqref{equ:cbf_merging}, \eqref{equ:sc_rear} and \eqref{equ:sc_merging} are conflict-free. Hence, the QP corresponding to time $t+\Delta t$ is feasible. \hfill $\blacksquare$

\begin{theorem}
\label{theorem:rear_merging2}
    Under \emph{Assumption \ref{assume:initial}} and \emph{Assumption \ref{assume:initial2}}, if $v_i \geq 0, u_{\min} \leq 0$, the QP \eqref{equ:ocbf_qp} subject to \eqref{equ:cbf1}, \eqref{equ:cbf_merging}, \eqref{VehicleConstraints}, \eqref{equ:sc_rear} and \eqref{equ:sc_merging} corresponding to any time interval $[t_{i}^{0} + k\Delta t, t_{i}^{0} + (k+1)\Delta t] \subset [t_i^0, t_i^m]$ is feasible.
\end{theorem}

\textbf{Proof:} \emph{Theorem} \ref{theorem:rear_merging2} is proved by induction following the same steps of the proof of \emph{Theorem} \ref{theorem:rear2} using \emph{Theorem \ref{theorem:rear_merging1}}. \hfill$\blacksquare$

\section{Simulation Results}

In this section, simulations are conducted to validate and assess the effect of the feasibility constraints added onto the OCBF framework. All simulations are performed in MATLAB using quadprog as the solver for the QPs.

We first build the model shown in Fig. \ref{modelF} with simulation parameters $L=400, u_{\min} = -2m/s^s, u_{\max} = 3m/s^2$ and adopt the OCBF controller without feasibility guarantee for each CAV. In some situations, a QP for optimally tracking the unconstrained optimal control trajectory of a CAV $i$ becomes infeasible. We record the indices of such CAVs and consider two different cases corresponding to the rear-end safety constraint and the safe merging constraint separately. We re-run the simulations of the two cases with feasibility constraints added to the ego CAV, keeping all other conditions same. The simulation results are detailed in what follows.

\subsection{Rear-end Safety Constraint}

\begin{figure}[ht]
    \centering
    \includegraphics[width = 0.9\linewidth]{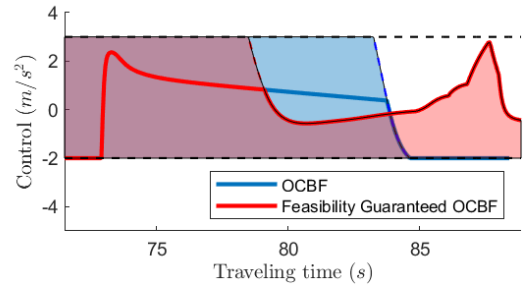}
    \caption{Control History Comparison}
    \label{fig:control1}
\end{figure}

A particular CAV, labeled ``CAV 25'', is chosen as the first case study. As CAV 25 and CAV 24 are in the same road, the possibly active constraint of interest is the rear-end safety constraint. We adopt the OCBF controller and run the simulation twice to derive the two trajectories of CAV 25, one with the feasibility guarantee and the other without. Note that in the merging problem, the control $u_i(t)$ is 1-dimensional, thus the feasible set of the QP is an interval. To illustrate the performance of the feasibility constraint, the evolution of the feasible set of the QPs over time are plotted in Fig. \ref{fig:control1}. 

In Fig. \ref{fig:control1}, the solid blue curve shows the control history $u(t)$ generated by the OCBF controller. The shaded blue area shows the feasible set of the QP. For each time $t$, the shaded blue area marks the maximum and minimum acceleration allowed by the QP. Note that when $t = 85s$, the QP becomes infeasible and the control is set to be $-2m/s^2$ to continue the program execution. The solid red line is the control history generated by the OCBF controller with the feasibility constraint added to the QP. The shaded red area shows the feasible set corresponding to the revised QP. The dashed black lines shows the control bounds of the CAV.

The figure shows that the OCBF and feasibility guaranteed OCBF have the same feasible set before 79s, which generates the same control history. However, after 79s, the feasible set of the feasibility guaranteed OCBF shrinks due to the feasibility constraint while the myopic OCBF approach keeps the same feasible set. This leads to a large difference after 84s. The feasible set of the myopic OCBF approach rapidly shrinks and even becomes empty, indicating that the QP is infeasible. The feasibility guaranteed OCBF, however, remains feasible with the help of the advance action introduced by the feasibility constraint. 

\begin{figure}[ht]
    \centering
    \includegraphics[width = 0.9\linewidth]{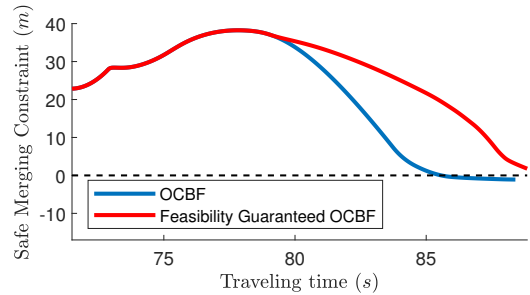}
    \caption{Rear-end Safety Constraint Satisfaction Comparison}
    \label{fig:re_sc}
\end{figure}

The infeasible QP makes the safety constraints unguaranteed as we no longer benefit from the forward invariance property of the CBF. The rear-end safety constraints of the two cases are plotted in Fig. \ref{fig:re_sc}, where the blue curve corresponds to the OCBF controller and the red curve corresponds to the feasibility guaranteed OCBF controller. From the figure, we can see that the rear-end safety constraint is violated after 85s using the OCBF controllers. This corresponds to the infeasible QP shown in Fig. \ref{fig:control1} after 85s. With the help of the feasibility constraint, the red curve remains positive, indicating that the rear-end safety constraint is always satisfied.

\subsection{Safe Merging Constraint}

\begin{figure}[ht]
    \centering
    \includegraphics[width = 0.9\linewidth]{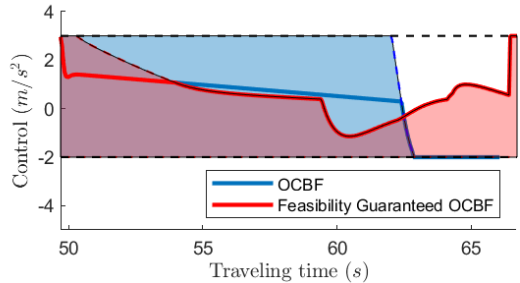}
    \caption{Control History Comparison}
    \label{fig:control2}
\end{figure}

\begin{figure}[ht]
    \centering
    \includegraphics[width = 0.9\linewidth]{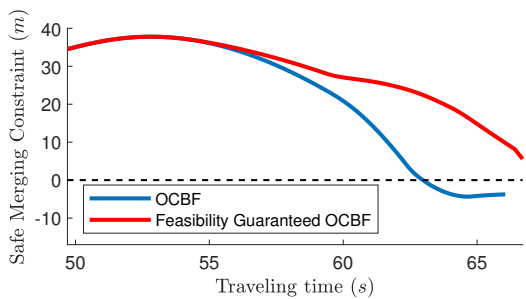}
    \caption{Safe Merging Constraint Satisfaction Comparison}
    \label{fig:safe_merging}
\end{figure}

The second case study focuses on CAV 17. As CAV 17 and CAV 16 are in different roads, the safe merging constraint needs to be considered. Following the same steps, we plot the control histories, as well as the safe merging constraints, of the two different approaches in Fig. \ref{fig:control2} and Fig. \ref{fig:safe_merging}. 


The simulation results are similar to the first case. The feasibility guaranteed OCBF is less myopic than the standard OCBF method and controls the CAV to decelerate earlier for safety, before it is impossible to do so. Unlike the rear-end safety constraint which is relatively easier to be satisfied (as shown in Fig. \ref{fig:re_sc}), the safe merging constraint is more likely to be violated. As is shown in Fig. \ref{fig:safe_merging}, when the QP becomes infeasible after 63s using OCBF, the safe merging constraint is violated and finally results in a potential car crash at 66s when merging. The feasibility guaranteed OCBF, however, avoids this problem with the feasibility constraint included.

\section{Conclusion}

We have presented a feasibility guaranteed decentralized control framework combining optimal control and CBFs leading to OCBF controllers for merging CAVs jointly minimizing both the travel time and the energy consumption subject to speed-dependent safety constraints, as well as vehicle limitations. We resolve the QP feasibility problem which arises when the OCBF controller is used in decentralized merging control by adding a single control-dependent feasibility constraint corresponding to each CBF constraint. Proofs are provided and the results are illustrated through simulations of a merging problem to demonstrate the effectiveness of this approach. The explicit control-dependent feasibility constraints we have derived rely on the special velocity-dependent safety constraint structure of the merging control problem. Thus, future research will explore the possibility of extending this method to deriving such feasibility constraints for arbitrary optimal control problems, as well as integrating recently developed event-driven QPs which relaxes the assumption of an adequately small time interval in \emph{Assumption \ref{assume:small_t}}.

\bibliographystyle{ieeetr}
\bibliography{cav}{}

\begin{thebibliography}{10}

\bibitem{rios2016survey}
J.~Rios-Torres and A.~A. Malikopoulos, ``A survey on the coordination of
  connected and automated vehicles at intersections and merging at highway
  on-ramps,'' {\em IEEE Transactions on Intelligent Transportation Systems},
  vol.~18, no.~5, pp.~1066--1077, 2016.

\bibitem{chen2015cooperative}
L.~Chen and C.~Englund, ``Cooperative intersection management: A survey,'' {\em
  IEEE Transactions on Intelligent Transportation Systems}, vol.~17, no.~2,
  pp.~570--586, 2015.

\bibitem{tideman2007review}
M.~Tideman, M.~C. van~der Voort, B.~van Arem, and F.~Tillema, ``A review of
  lateral driver support systems,'' in {\em 2007 IEEE Intelligent
  Transportation Systems Conference}, pp.~992--999, IEEE, 2007.

\bibitem{li2013survey}
L.~Li, D.~Wen, and D.~Yao, ``A survey of traffic control with vehicular
  communications,'' {\em IEEE Transactions on Intelligent Transportation
  Systems}, vol.~15, no.~1, pp.~425--432, 2013.

\bibitem{xu2019grouping}
H.~Xu, S.~Feng, Y.~Zhang, and L.~Li, ``A grouping-based cooperative driving
  strategy for cavs merging problems,'' {\em IEEE Transactions on Vehicular
  Technology}, vol.~68, no.~6, pp.~6125--6136, 2019.

\bibitem{xu2020bi}
H.~Xu, Y.~Zhang, C.~G. Cassandras, L.~Li, and S.~Feng, ``A bi-level cooperative
  driving strategy allowing lane changes,'' {\em Transportation research part
  C: emerging technologies}, vol.~120, p.~102773, 2020.

\bibitem{milanes2010automated}
V.~Milan{\'e}s, J.~Godoy, J.~Villagr{\'a}, and J.~P{\'e}rez, ``Automated
  on-ramp merging system for congested traffic situations,'' {\em IEEE
  Transactions on Intelligent Transportation Systems}, vol.~12, no.~2,
  pp.~500--508, 2010.

\bibitem{rios2015online}
J.~Rios-Torres, A.~Malikopoulos, and P.~Pisu, ``Online optimal control of
  connected vehicles for efficient traffic flow at merging roads,'' in {\em
  2015 IEEE 18th international conference on intelligent transportation
  systems}, pp.~2432--2437, IEEE, 2015.

\bibitem{bichiou2018developing}
Y.~Bichiou and H.~A. Rakha, ``Developing an optimal intersection control system
  for automated connected vehicles,'' {\em IEEE Transactions on Intelligent
  Transportation Systems}, vol.~20, no.~5, pp.~1908--1916, 2018.

\bibitem{hult2016coordination}
R.~Hult, G.~R. Campos, E.~Steinmetz, L.~Hammarstrand, P.~Falcone, and
  H.~Wymeersch, ``Coordination of cooperative autonomous vehicles: Toward safer
  and more efficient road transportation,'' {\em IEEE Signal Processing
  Magazine}, vol.~33, no.~6, pp.~74--84, 2016.

\bibitem{cao2015cooperative}
W.~Cao, M.~Mukai, T.~Kawabe, H.~Nishira, and N.~Fujiki, ``Cooperative vehicle
  path generation during merging using model predictive control with real-time
  optimization,'' {\em Control Engineering Practice}, vol.~34, pp.~98--105,
  2015.

\bibitem{mukai2017model}
M.~Mukai, H.~Natori, and M.~Fujita, ``Model predictive control with a mixed
  integer programming for merging path generation on motor way,'' in {\em 2017
  IEEE Conference on Control Technology and Applications (CCTA)},
  pp.~2214--2219, IEEE, 2017.

\bibitem{nor2018merging}
M.~H. B.~M. Nor and T.~Namerikawa, ``Merging of connected and automated
  vehicles at roundabout using model predictive control,'' in {\em 2018 57th
  Annual Conference of the Society of Instrument and Control Engineers of Japan
  (SICE)}, pp.~272--277, IEEE, 2018.

\bibitem{xiao2020bridging}
W.~Xiao, C.~G. Cassandras, and C.~Belta, ``Bridging the gap between optimal
  trajectory planning and safety-critical control with applications to
  autonomous vehicles,'' {\em Automatica}, vol.~129, p.~109592, 2021.

\bibitem{Xiao2019}
W.~Xiao and C.~Belta, ``Control barrier functions for systems with high
  relative degree,'' in {\em Proc. of 58th IEEE Conference on Decision and
  Control}, (Nice, France), pp.~474--479, 2019.

\bibitem{xiao2021decentralized}
W.~Xiao and C.~G. Cassandras, ``Decentralized optimal merging control for
  connected and automated vehicles with safety constraint guarantees,'' {\em
  Automatica}, vol.~123, p.~109333, 2021.

\bibitem{Xiao2021}
W.~Xiao, C.~Belta, and C.~G. Cassandras, ``Sufficient conditions for
  feasibility of optimal control problems using control barrier functions,''
  {\em Automatica (in print), preprint arXiv:2011.08248}, 2021.

\bibitem{XiaoITSC2020}
W.~Xiao, C.~G. Cassandras, and C.~Belta, ``Decentralized optimal control in
  multi-lane merging for connected and automated vehicles,'' in {\em 2020 IEEE
  23rd International Conference on Intelligent Transportation Systems (ITSC)},
  pp.~1--6, 2020.

\bibitem{XiaoACC2020}
W.~Xiao and C.~G. Cassandras, ``Decentralized optimal merging control for
  connected and automated vehicles with optimal dynamic resequencing,'' in {\em
  2020 American Control Conference (ACC)}, pp.~4090--4095, 2020.

\bibitem{Vogel2003}
K.~Vogel, ``A comparison of headway and time to collision as safety
  indicators,'' {\em Accident Analysis \& Prevention}, vol.~35, no.~3,
  pp.~427--433, 2003.

\bibitem{Kamal2013}
M.~Kamal, M.~Mukai, J.~Murata, and T.~Kawabe, ``Model predictive control of
  vehicles on urban roads for improved fuel economy,'' {\em IEEE Transactions
  on Control Systems Technology}, vol.~21, no.~3, pp.~831--841, 2013.

\bibitem{Bryson1969}
Bryson and Ho, {\em Applied Optimal Control}.
\newblock Waltham, MA: Ginn Blaisdell, 1969.

\bibitem{Xiao2021D}
W.~Xiao, C.~Belta, and C.~G. Cassandras, ``Event-triggered safety-critical
  control for systems with unknown dynamics,'' in {\em Proc. of 60th IEEE
  Conference on Decision and Control, preprint in arXiv:2103.15874}, 2021.

\bibitem{zhangACC2017}
Y.~Zhang, C.~G. Cassandras, and A.~A. Malikopoulos, ``Optimal control of
  connected automated vehicles at urban traffic intersections: A feasibility
  enforcement analysis,'' in {\em 2017 American Control Conference (ACC)},
  pp.~3548--3553, 2017.

\end{thebibliography}

\end{document}